\documentclass[a4paper,12pt]{article}
\def\al{\alpha}
\def\be{\beta}
\def\ga{\gamma} \def\Ga{\Gamma}
\def\ep{\epsilon}
\def\lam{\lambda}
\def\Lam{\Lambda}

\def\calD{{\cal D}}  
  
  \def\calL{{\cal L}}
 \def\calN{{\cal N}} \def\calO{{\cal O}}

\def\del        {  \partial  }
\def\half       {  {1\over 2}  }
\def\trace      {  \mbox{Tr}  }

\def\ie         {  {\it i.e.}      }
\def\where      { \mbox{where}\qquad }
\def\comma          {\, ,}
\def\period         {\, .}
\def\lsim    {\lower .65ex \hbox{\ $\stackrel{<}{\sim}$\ } }
\def\gsim    {\lower .65ex \hbox{\ $\stackrel{>}{\sim}$\ } }
\def\com#1#2   { \left[#1, #2\right]} 
\def\acom#1#2  {\left\{ #1,#2\right\}}
\def\bra#1     {\langle #1 |}
\def\ket#1     {| #1 \rangle}
\def\vecii#1#2      {  \left(\begin{array}{c}#1\\#2\end{array}\right)  }
\def\veciii#1#2#3   {  \left(\begin{array}{c}#1\\#2\\#3\end{array}
                     \right)  }
\def\veciv#1#2#3#4  {  \left(\begin{array}{c}#1\\#2\\#3\\#4
                                 \end{array}\right)  }
\def\vecfv#1#2#3#4#5 {  \left(\begin{array}{c}#1\\#2\\#3\\#4\\#5
                                 \end{array}\right)  }

\def\matrixii#1#2#3#4            {  \left(\begin{array}{cc}#1&#2\\#3&#4
                                       \end{array}\right) }
\def\matrixiii#1#2#3#4#5#6#7#8#9 {  \left(\begin{array}{ccc}#1&#2&#3\\
                                     #4&#5&#6\\#7&#8&#9\end{array}
                               \right)  }
\def\mativ#1#2#3#4               {  \left(\begin{array}{cccc}
                                       #1\\#2\\#3\\#4\end{array}\right) }
\def\matv#1#2#3#4#5              {  \left(\begin{array}{ccccc}
                                     #1\\#2\\#3\\#4\\#5\end{array}
                              \right)  }
\def\eqabegin         {  \begin{eqnarray}  }
\def\eqaend           {  \end{eqnarray}  }
\def\nn               {  \nonumber  }
\def\bracetwo#1#2     {  \left\{ \begin{array}{l} #1 \\ #2 \end{array}
                         \right.  }
\def\bracetwocases#1#2#3#4  {   \left\{ \begin{array}{ll} #1 &
                                 \qquad #2 \\
                                 #3 & \qquad #4 \end{array} \right.  }
\def\bracebegin#1     {  \left\{ \begin{array}{#1}   }
\def\braceend         {  \end{array}\right.   }
\def\parn              {  \par\noindent }

\def\parmedskip        {  \par\medskip  }
\def\parsmallskip      {  \par\smallskip  }
\def\parbigskipn        {  \par\bigskip\noindent  }

\def\parsmallskipn      {  \par\smallskip\noindent  }

\def\parag#1           {\paragraph{#1} \mbox{ }\parmedskip\noindent}
\def\msection#1      {  \begin{center} \section{#1} \end{center}   }
\def\nsection#1      {  \let\boldface\bf \def\bf{} \section{#1}
                           \let\bf\boldface   }
\def\mnsection#1     {  \begin{center} \nsection{#1} \end{center}  }
\def\capsection#1    {  \let\boldface\bf \def\bf{\sc} \section{#1}
                           \let\bf\boldface   }
\def\mcapsection#1   {  \begin{center} \capsection{#1} \end{center} }

\def\sectionnumbering { \setcounter{equation}{0}
         \renewcommand{\theequation}{\arabic{section}.\arabic{equation}}}

\newcommand{\nullify}[1]{}
\def\papertitlepage{\baselineskip 3.5ex \thispagestyle{empty}}
\def\Title#1{\baselineskip 1cm \vspace{1.5cm}\begin{center}
 {\Large\bf #1} \end{center} 
\vspace{0.5cm}}
\def\Authors#1{\begin{center} {\it #1} \end{center}}
\def\Abstract{\vspace{1.0cm}\begin{center} {\large\bf Abstract} 
           \end{center} \par\bigskip}
\def\Komabanumber#1#2#3{\hfill \begin{minipage}{4.2cm} UT-Komaba #1
              \parn #2 
              \parn #3 \end{minipage}}
\renewcommand{\thefootnote}{\fnsymbol{footnote}}
\renewenvironment{thebibliography}{\pagebreak[3]\par\vspace{0.6em}
\begin{flushleft}{\large \bf References}\end{flushleft}
\vspace{-1.0em}

\begin{enumerate}\if@twocolumn\baselineskip=0.6em\itemsep -0.2em
\else\itemsep -0.2em\fi\labelsep 0.1em}{\end{enumerate}}

\def\Btil{\widetilde{B}}
\def\Gatil{\widetilde{\Ga}}

\def\Cbar{\overline{C}}

\def\Yhat{\widehat{Y}}

\def\Atil{\widetilde{A}}

\def\Wtil{\tilde{W}}

\def\rslash{\slash{r} }
\def\vslash{\slash{v} }

\def\thdot{\dot{\theta}}
\def\vr{v\cdot r}
\def\Stil{\widetilde{S}}

\def\Gtil{\widetilde{G}}
\def\Yhat{\widehat{Y}}
\def\Psihat{\hat{\Psi}}
\def\Ztil{\widetilde{Z}}
\def\Wtil{\widetilde{W}}
\def\yhat{\widehat{y}}
\def\psihat{\widehat{\psi}}
\def\slash#1{#1\!\!\!/}

\def\Ebar{\bar{E}}
\def\Fbar{\bar{F}}

\def\tilA{\widetilde{A}}
\def\lk{\langle}
\def\rk{\rangle}
\def\taup{\tau'}
\def\taupp{\tau''}
\usepackage{graphicx}
\renewenvironment{thebibliography}{\pagebreak[3]\par\vspace{0.6em}
\begin{flushleft}{\large \bf References}\end{flushleft}
\vspace{-1.0em}

\begin{enumerate}\if@twocolumn\baselineskip=0.6em\itemsep -0.2em
\else\itemsep -0.2em\fi\labelsep 0.1em}{\end{enumerate}}
\begin{document}
\papertitlepage
\vspace*{0cm}
\Komabanumber{00-05}{hep-th/0003161}{March, 2000}
\Title{On the Supersymmetry and Gauge Structure of Matrix Theory} 
\vspace{1cm}
\Authors{{\sc Y.~Kazama\footnote[2]{kazama@hep3.c.u-tokyo.ac.jp} 
 and T.~Muramatsu
\footnote[3]{tetsu@hep1.c.u-tokyo.ac.jp}
\\ }
\vskip 3ex
 Institute of Physics, University of Tokyo, \\
 Komaba, Meguro-ku, Tokyo 153-8902 Japan \\
  }
\baselineskip .7cm
\Abstract
Supersymmetric Ward identity for the low energy effective action 
in the standard background gauge is derived for {\it arbitrary}
trajectories of supergravitons in  Matrix Theory.  
In our formalism, the quantum-corrected supersymmetry transformation 
laws of the supergravitons are directly identified in  closed form, 
which exhibit an intricate interplay between supersymmetry and gauge
(BRST) symmetry. As an application, we explicitly compute the transformation 
laws for the source-probe configuration at 1-loop and 
confirm that supersymmetry fixes the form of the action completely,
including the normalization, to the lowest order in 
the derivative expansion.
 
\newpage
\baselineskip 3.3ex
\section{Introduction}  
 \sectionnumbering
\renewcommand{\thefootnote}{\arabic{footnote}}
By now a considerable amount of evidence has been 
accumulated for the Matrix theory for M-theory, 
originally proposed by 
Banks, Fischler, Shenker and Susskind\cite{bfss} and later re-interpreted 
by Susskind\cite{susskind} in the framework of 
discrete light-cone quantization. 
In particular, just to mention only the direct comparison 
with eleven dimensional 
 supergravity,  complete agreement for the multi-graviton scattering
 (including the recoil effects) at 2-loop\cite{Okawa-Yoneya,Okawa-Yoneya2}
 and 
 that for the two-body potential between arbitrary fermionic as well as 
 bosonic objects at 1-loop
\cite{TR} can be cited as highly non-trivial and remarkable. 

Despite such impressive pieces of evidence as well as general supportive 
arguments
\cite{sen,seiberg}, the deep reason and 
the mechanism of the agreement are yet to be fully understood. 
Evidently, one of the keys should be the understanding of the structure and 
the role of the symmetries. The well-known symmetries present both in 
 the Matrix theory and in the supergravity theory 
are the global $Spin(9)$ invariance, the CPT invariance and  
 the invariance under 16 supersymmetries.
 Perhaps less familiar  is the generalized conformal 
invariance\cite{jy,jky,jky2},
  the generalization of the conformal symmetry that plays the major role in 
 the AdS/CFT correspondence of Maldacena\cite{maldacena}. Besides 
 these symmetries, the supergravity possesses the general coordinate
 invariance while the Matrix theory has the Yang-Mills type gauge symmetry, 
 which must be deeply connected. Finally, although not yet 
 clearly identified, the agreement of the multi-body scattering 
 amplitudes strongly suggests that the eleven dimensional Lorentz invariance
 is present in a highly non-trivial manner in the Matrix theory. 

Except for the eleven dimensional  Lorentz invariance, 
the above-mentioned symmetries of the Matrix theory are easily 
 recognized in the original action. 
However, since the supergravity interactions between various objects 
 arise {\it only after} introducing the corresponding backgrounds 
and integrating out the quantum fluctuations around them, the 
 realizations of these symmetries are in general modified 
 for the effective action of interest. Besides, being an off-shell 
 quantity, the form of the effective action depends on the choice of 
 the gauge as well as on the definition of the fields.  Further, 
there is always a vast degree of ambiguities as one can add total derivatives.
For these reasons, the study of how the symmetries govern the structure 
 of the effective action becomes quite non-trivial. 

In this article, we shall focus on the supersymmetry (SUSY),
 considered to be the most powerful among the ones listed above. In fact 
it has been claimed that a large degree of SUSY, $\calN=16$, 
present in the theory imposes  strong restrictions on the form 
 of the effective action and leads to a number of \lq\lq non-renormalization
 theorems"\cite{Pabanetal1}--\nocite{Pabanetal2}
\nocite{lowe}\nocite{ss}\nocite{Hyunetal}\cite{np}. 
However, upon close examinations one finds that 
 such assertions in the existing literature can still be challenged.
 This is essentially due to the lack of completely off-shell 
 consideration. As we shall elaborate in some detail in Sec.~3, 
to understand precisely to what extent SUSY is responsible in determining 
 the effective action, one must allow the background fields to have 
  {\it arbitrary}
 time-dependence. This in turn inevitably leads to the necessity
 of examining the gauge (BRST) symmetry, another important ingredient 
 of Matrix theory, as SUSY and  gauge symmetry are known to be 
intimately intertwined off-shell. Such an analysis has not been 
performed in the past. 

What makes the off-shell analysis difficult is that we do not as yet 
 have the off-shell unconstrained superfield formulation in the case of 
 $\calN=16$ supersymmetry. This prompts us to resort to the conventional 
means, namely the Ward identity for the symmetries in question. 
In order to be able to check against the available explicit result for 
  the effective action, one would like to obtain the Ward identity 
 in so-called the \lq\lq background gauge", in which all the calculations
 have been performed. Although the derivation of such a  Ward identity
 is expected to be a text-book matter, this was not to be: 
One must carefully disentangle the dependence on the background field 
of the effective action by making use of BRST Ward identities. 
The result is a somewhat complicated Ward identity, in which 
 the supersymmetry and the BRST symmetry are intertwined. A notable
 feature of our Ward identity is that one can read off the effective 
quantum-corrected  SUSY transformations in  closed form and this should 
 serve as a starting point of various truly off-shell investigations. 
As a simple application, we  compute the transformation 
 laws to the lowest order in the derivative expansion at 1-loop and 
 analyze the restriction imposed by SUSY on the effective action for 
 a background with arbitrary time-dependence, to the corresponding order.  
We find that at this order the effective action is indeed fully determined 
 by the requirement of supersymmetry, which agrees with 
 the explicit calculation\cite{okawa9907} {\it including} the  normalization. 

The rest of the paper is organized as follows: In Sec.~2,  we 
 recall the  symmetries of the original action of Matrix theory 
and the BRST symmetry associated with the background gauge fixing. 
Then in Sec.~3  we make some important remarks on the determination of 
 the effective action and its symmetries, to emphasize the necessity 
 of completely off-shell analysis.  Having clarified the issue, 
 we proceed in Sec.~4 to the derivation of the SUSY Ward identity 
 for the effective action in the background gauge.  By carefully 
 separating the different origins of the dependence on the background 
 field, we obtain the desired Ward identity together with the 
 closed expressions for the effective SUSY transformation laws. This 
 constitutes the main result of this work. As an application, 
 we explicitly work out in Sec.~5 the SUSY transformation laws to the lowest 
 order in the derivative expansion at 1-loop and show that the effective 
action to the corresponding order is completely determined by the 
 Ward identity. Sec.~6 is devoted to a short summary and discussions of 
 future problems. 
\section{Action and its Symmetries}
Let us begin by recalling the action of the Matrix theory and its 
 symmetries, which at the same time serves to set our notations. 

The basic action of the Matrix theory can be written as
\begin{eqnarray}
S_0 &=&
\trace \int\! dt \Biggl\{ \half  \com{D_t}{X_m} ^2 + {g^2\over 4}
 \com{X_m}{X_n} ^2  \nn\\
&& \qquad +{i\over 2} \Theta^T \com{D_t}{\Theta} + {g\over 2}
\Theta^T \ga^m\com{X_m}{\Theta} \Biggr\}\comma  \\
D_t &=& \del_t -igA \comma 
\end{eqnarray}
where $N\times N$  hermitian matrices $X_{ij}^m(t)\comma  A_{ij}(t)
\comma  \Theta_{\al,ij}(t)$ stand for the bosonic, the gauge, and the 
 fermionic fields respectively. The middle Latin indices $m,n, \ldots$, 
running from $1$ to $9$, denote the spatial directions, while the 
Greek letters such as  $\al\comma \be=1\sim 16$ are used for the $SO(9)$ 
spinor indices. The $16\times 16$ $\ga$-matrices $\ga^m$ are real symmetric 
 and satisfy $\acom{\ga^m}{\ga^n} =\delta^{mn}$. 

To facilitate the quantum computations, it is convenient to define the theory 
 by going to the \lq\lq Euclidean formulation". Introduce the Euclidean 
 time $\tau$, the gauge field $\Atil$,  and the action $\Stil_0$ 
 by\footnote{Fermions are not transformed.}
\begin{eqnarray}
\tau &\equiv & it\comma \qquad \Atil \equiv -iA\comma \qquad 
 \Stil_0 \equiv -iS_0 \period
\end{eqnarray}
Then the action and the covariant derivative become
\begin{eqnarray}
\Stil_0 &=& \trace \int\! d\tau \Biggl\{ \half \com{D_\tau}{X_m} ^2 
- {g^2\over 4}
 \com{X_m}{X_n} ^2  \nn\\
&& \qquad +\half\Theta^T \com{D_\tau}{\Theta} 
-\half g \Theta^T \ga^m\com{X_m}{\Theta} \Biggr\}\comma \\
D_\tau &=& \del_\tau -ig\Atil\period
\end{eqnarray}
Besides the obvious $Spin(9)$ symmetry, this action is invariant
 under the following transformations:
\begin{enumerate}
	\item Gauge transformations with a gauge parameter matrix $\Lam$:
\begin{eqnarray}
\delta_\Lam \Atil &=& \com{D_\tau}{\Lam} \comma \qquad 
\delta_\Lam X_m = ig\com{\Lam}{X_m} \comma \qquad 
\delta_\Lam \Theta = ig\com{\Lam}{\Theta} \period
\end{eqnarray}
	\item Supersymmetry transformations with a spinor parameter $\ep_\al $:
\begin{eqnarray}
\delta_\ep \Atil &=& \ep^T \Theta \comma \qquad 
\delta_\ep X^m = -i\ep^T \ga^m\Theta \comma \label{susyb}\\
\delta_\ep \Theta &=& i\left( \com{D_\tau}{X_m} \ga^m
 +{g\over 2} \com{X_m}{X_n} \ga^{mn} \right) \ep\label{susyf}
\end{eqnarray}
where $\ga^{mn} \equiv  \half \com{\ga^m}{\ga^n} $ is real anti-symmetric. 
When the system possesses symmetries other than the supersymmetry, the 
supersymmetry algebra may close not only on the space-time translation but
 also on the generators of such additional symmetries. 
In fact in the present case it is well-known that it involves the gauge
 symmetry with a field-dependent gauge function.  For example on $\Atil$, 
\begin{eqnarray}
\com{\delta_\ep}{\delta_\lam} \Atil &=& -2i \ep^T \lam \, \del_\tau \Atil
 + \delta_\Lam \Atil \comma \\
\where \Lam &=& -2i(\ep^T\ga^m \lam X_m -\ep^T\lam \Atil)\comma 
\end{eqnarray}
and similarly for the other fields. Moreover, for a system with 
$\calN=16$ supersymmetry, such as the Matrix theory, formulation in 
 terms of unconstrained superfields is not known and hence the algebra
 closes only up to the equations of motion in general. 
Thus it is expected that the proper understanding of the supersymmetry 
 of the Matrix theory must include the analysis of these non-trivial features.

 \item Generalized conformal transformations: If we rescale the fields 
$X_m$ and $\Atil$  by a factor of $g$, such as $X_m \rightarrow X_m/g$, 
and allow $g$  to depend on $\tau$ to a linear order, $\Stil_0$ is 
invariant under a generalization  of the conformal transformations\cite{jy}
\cite{jky2}.
 In particular, the invariance under the special conformal transformation 
defined by
\begin{eqnarray}
\delta_K \Atil &=& 2\ep \tau \Atil\comma\qquad 
 \delta_K X_m = 2\ep \tau X_m \comma \qquad \delta_K \Theta_\al =0 \nn\\
\delta_K \tau &=& -\ep \tau^2\comma \qquad \delta_K g = 3\ep \tau g\comma \\
\ep &=& \mbox{an infinitesimal bosonic parameter}\comma \nn
\end{eqnarray}
imposes a useful restriction on the form of the effective action. 
\end{enumerate}

Besides these well-established symmetries, the remarkable agreement 
between the 11-dimensional supergravity calculations
 and the 2-loop Matrix theory calculations for 
 multi-body scattering processes\cite{Okawa-Yoneya}
 strongly suggests that the Matrix theory
 actually possesses 11-dimensional Lorentz symmetry in a highly non-trivial
 manner.
\parmedskip
In this article, we shall focus on how the first two of these symmetries, 
 which  are intimately intertwined, are implemented in 
 the quantum effective action of the supergravitons. In the M-theory 
 interpretation of the Matrix theory, the coordinates and the 
 spin degrees of freedom  of these supergravitons
 are represented by the diagonal backgrounds  for $X^m$ and $\Theta_\al$
 respectively.  We shall denote them by $B_m$ and $\theta_\al$ 
 respectively and separate them 
  from the quantum parts $Y_m$ and $\Psi_\al$  as 
\begin{eqnarray}
X_m &=& {1\over g}B_m + Y_m\comma \\
\Theta_\al &=& {1\over g}\theta_\al + \Psi_\al \period
\end{eqnarray}
As was already emphasized in the introduction and will be further 
 elaborated  in the next section,  it is important to take
 $B_m(\tau)$ and $\theta_\al(\tau)$ as {\it arbitrary} backgrounds,
 not satisfying any equations
 of motion. Only in this way  we can unambiguously determine how much 
restrictions are imposed by the supersymmetry on the effective action 
for these background fields.  

To quantize the theory, we need to fix the gauge. Although our derivation 
 of the Ward identity, to be presented in Sec.~4, 
 can be readily adapted to any choice of gauge, the actual computations 
 are extremely cumbersome except in the {\it standard  background 
 gauge}. It is  specified by the gauge-fixing function of the form
\begin{eqnarray}
G &=& -\del_\tau \Atil + i\com{B^m}{X_m} \period \label{sbgg}
\end{eqnarray}
In fact essentially all the existing explicit calculations have been 
performed in  this gauge. However, as it will become clear, 
 the naive use of this gauge leads to a subtle but important 
complication in deriving 
 the correct Ward identity. To avoid this problem, we will tentatively 
 use a {\it different} function $\Btil_m$ in place of $B_m$ and write the 
 gauge-fixing function as 
\begin{eqnarray}
\Gtil &=& -\del_\tau \Atil + i\com{\Btil^m}{X_m} \period
\end{eqnarray}
Later at an appropriate stage, we will set $\Btil_m=B_m$. 

The corresponding ghost action can be readily obtained by the standard 
 BRST method. The BRST transformations for the quantum part of the 
 fields are given by
\begin{eqnarray}
\delta_B \Atil &=&  \com{D_\tau}{C} \comma \qquad 
\delta_B Y_m = -ig\com{X_m}{C} \comma \nn\\
\delta_B \Psi &=& ig\acom{C}{\Theta} \comma \label{brstr}\\
\delta_B C &=& igC^2 \comma \qquad \delta_B \Cbar = ib
\comma \qquad \delta_B b=0\period \nn
\end{eqnarray}
$C$, $\Cbar$ and $b$ are, respectively, the ghost, the anti-ghost and 
 the Nakanishi-Lautrup auxiliary fields. The background fields 
 are {\it not} transformed. Then the combined gauge-ghost action $\Stil_{gg}$
 is  generated by 
\begin{eqnarray}
\Stil_{gg} &=& \delta_B \trace \int\! d\tau \left[ {1\over i} \Cbar \left(\Gtil-
{\al \over 2} b\right) \right]\period \label{sgg}
\end{eqnarray}
We will henceforth set the gauge parameter $\al$ to be 1. This leads,
 after integrating over the $b$ field, to the familiar gauge-ghost action
\begin{eqnarray}
\Stil_{gg} &=& \trace \int\! d\tau \half \left(-\del_\tau \Atil 
+ i\com{\Btil^m}{X_m} \right)^2  \nn\\
&& \qquad -i \trace \int\! d\tau \left( \Cbar \del_\tau \com{D_\tau}{C} 
 -g\Cbar \com{\Btil^m}{\com{X_m}{C} } \right)\period
\end{eqnarray}
\section{Remarks on the Determination of Effective Action and its
 Symmetries}
Before starting the derivation of the off-shell supersymmetry Ward identity, 
 we wish to make some important remarks on the determination of the effective
 action and its symmetries, which point to  the necessity of off-shell 
 analysis. Although many of the remarks will apply for general 
 backgrounds, for clarity of discussions we shall consider the so-called 
 \lq\lq source-probe configuration".  In the case of $U(N+1)$ gauge group, 
 it is defined as the situation where a probe supergraviton interacts with 
 $N$ supergravitons situated at the origin that act as a heavy source.
The background fields representing this situation are 
\begin{eqnarray}
B_m &=& {\rm diag}\, (r_m, 0,0, \ldots , 0) \comma \qquad 
 \theta_\al = {\rm diag}\, (\theta_\al,0,0,\ldots, 0) \period
\label{sourceprobe}
\end{eqnarray}
Here, 
$r_m$ represents the coordinate of the probe and $\theta_\al$ its 
 spin content. As usual, the spin of the source is neglected. 

As already emphasized in the introduction, the primary feature of Matrix 
 theory is that it generates the supergravity interactions among various 
 objects {\it only after} (i)\ introducing the corresponding backgrounds 
 and (ii)\ integrating over the quantum fluctuations around them. 
Because of this,  realizations of the symmetries of the effective theory 
 are in general modified non-trivially. 
Besides, being an off-shell quantity, the form of the effective action is 
 affected by  (A)\ the gauge choice, (B)\ the (re)definition of the fields, 
and  (C)\ the freedom of adding total derivatives. Of course the on-shell
 S-matrix elements do not depend on these factors. However, the determination  
of the {\it full} (\ie quantum-corrected) on-shell condition itself requires 
the knowledge of the off-shell effective action\footnote{This was 
 clearly demonstrated in \cite{okawa9907},  where the agreement between 
the supergravity and the Matrix theory calculation was achieved with 
 the recoil corrections}. Thus, in order to 
 understand the symmetry structure of the effective theory fully, 
it is necessary to perform an off-shell analysis with (A) $\sim$ (C) 
properly taken into account. 
 
Now since an exact analysis is practically impossible, one often needs to make 
 some approximations. In doing so, one must make sure that they 
 are logically consistent for one's aim. For the present purpose, 
 some of the often used approximations are not appropriate. 
For example, the eikonal approximation, where one tries to reconstruct
 the effective action from the eikonal phase shift, can be dangerous and 
 misleading. In fact the answer depends on 
 the form of the effective Lagrangian assumed. As a simple illustration, 
 consider the 1-loop eikonal phase shift \cite{dkps} for $v\ll 1$ given by 
 \begin{eqnarray}
\Gatil_1^e &=& -{v^3 \over b^6} + 0\times {v^5 \over b^{10}} 
-{3\over 2}{v^7 \over b^{14} }+ \calO(v^9) \nn\\
r_m &=& v_m\tau + b_m\comma \quad v\cdot b=0\comma 
\qquad b_m=\mbox{impact parameter} \period \nn
\end{eqnarray}
If one assumes the effective Lagrangian to be of the form 
  $\calL = \calL(v,r)$, then the effective action that reproduces 
 this phase shift is uniquely determined to be 
\begin{eqnarray}
\Stil_1 &=& \int\! d\tau\left(-{15\over 16}{v^4\over r^7} 
+ 0\times {v^6\over r^{11}}
 -{9009\over 4096}{v^8\over r^{15}} + \calO(v^{10})\right) \nn\period
\end{eqnarray}
However, even restricting to $\calO(v^4)$, the most general 
form allowed for the effective action contains 6 independent structures,
 after eliminating total-derivative ambiguities:
\begin{eqnarray}
\Stil_1 &=& \int\! d\tau \left( A{v^4\over r^7} + B{v^2 (\vr)^2 \over r^9}
 + C{(\vr)^4 \over r^{11}}\right. \nn\\
&& \left. + D{v^2 (a\cdot r) \over r^7} + E{(a\cdot r)^2 \over r^7}
 + F{a^2\over r^5}\right)\period \nn
\end{eqnarray}
On the other hand, there is only one condition, 
$-(15/16) = A + (1/7) B + (1/21) C$, required for the correct 
 phase shift, and hence 5 parameters remain undetermined. 
The situation at $\calO(v^6)$ is even more striking. Although 
 the eikonal phase shift vanishes, the explicit computation reveals that
 there are 5 non-vanishing independent structures present in the effective 
 action\cite{okawa9903}. 

It should be clear from these illustrations that the only logically 
 consistent procedure, not affected by the total-derivative ambiguities, 
 is to use off-shell backgrounds with {\it arbitrary} 
$\tau$-dependence\footnote{This was emphasized in the context of 
 generalized conformal symmetry in \cite{hm9901}. Related discussion can 
 also be found in \cite{okawa9907}.} and to classify terms by derivative 
expansion according to the \lq\lq order" defined by 
\begin{eqnarray}
\mbox{order} &=& \mbox{\# of $\del_\tau$} + \half \mbox{\# of fermions}
\period 
\label{deforder}
\end{eqnarray}
If necessary, one may combine this with the usual loop expansion. 

Having emphasized the importance of off-shell considerations, 
we now make some related comments on the general arguments on the 
restrictions imposed by supersymmetry, often referred to as 
 SUSY non-renormalization theorems.  They can be roughly classified into 
 two categories. 

The first type of argument, devised by Paban et al \cite{Pabanetal1},
 relies on the closure property of SUSY transformations. 
For example, at $\calO(v^2)$  they first make a choice of the definition 
of the fields so that the action takes the form
$\displaystyle\int\! d\tau f(r) v^2$ and 
 take the $\calO(v^0)$ SUSY transformation laws {\it in that basis}
 to be the standard 
 ones without any correction, $\delta_\ep r^m =
 -i\ep \ga^m \theta\comma \delta_\ep \theta_\al = i(\vslash \ep)_\al$. 
Then demanding that the closure is canonical, namely 
 $\com{\delta_\ep}{\delta_\lam} = 2\lam^T\ep \del_\tau$, they show that 
 there cannot be a correction to the transformation laws 
at $\calO(v^2)$ and hence the  $\calO(v^2)$ effective action is tree-exact. 
Although the argument is quite simple and plausible, 
it is unclear why the closure should 
 be canonical {\it off-shell} and further it is not obvious if the $\calO(v^0)$
 transformation  laws must be of the standard form in a particular basis 
 adopted. In general, field redefinitions affect  the form of the 
 SUSY transformations and hence they must be considered as a pair. 

The other type of argument is known as the SUSY completion method, 
 which makes use of the chain of relations produced among terms with 
 different number of $\theta$'s by  SUSY transformations. For example, 
 at $\calO(v^4)$ one expects  relations of the form 
\begin{eqnarray}
v^4 \longleftarrow v^3 \theta^2 \longleftarrow v^2 \theta^4 \longleftarrow
 v \theta^6 \longleftarrow \theta^8 \period \label{v4chain}
\end{eqnarray}
By showing that the top form, $\theta^8$ term in this case, is not
 renormalized beyond 1-loop, one wishes to infer the non-renormalization
 of all the other terms in the chain, in particular the   $v^4$ term.
This method appears efficient, but some care is needed in drawing  
 firm conclusions. One problem is that sometimes the chain starting from 
 the top form stops at an intermediate stage. Put differently, 
one may form a super-invariant not containing the top form. An example already 
 occurs at $\calO(v^2)$, where the tree-level expression 
$\displaystyle\int\! d\tau \left((v^2/2g^2) +
 (\theta\thdot/2g^2)\right)$
is SUSY-complete without
 a $\theta^4$ term. More non-trivial example is seen at $\calO(v^6)$: Although
 $\theta^{12}$ term was shown to vanish\cite{Pabanetal2} at 1-loop, 
the bosonic contribution at $\calO(v^6)$ nonetheless exists\cite{okawa9903}. 

An attempt at filling this gap was made in \cite{Hyunetal}. In this work 
all the connections in the chain (\ref{v4chain}) were examined and it was
 concluded that SUSY is indeed powerful enough to fix the effective action
 at this order up to an overall constant. Again one must be cautious in 
 accepting this conclusion: In this analysis $v$ and $\theta$ were taken 
 to be $\tau$-independent and hence the assumed form of the effective
 Lagrangian was not the most general one allowed in the proper derivative 
expansion with arbitrary backgrounds. Later it was recognized\cite{okawa9907},
however, that the higher derivative terms neglected in this analysis can 
actually be absorbed into the tree-level Lagrangian by a 
 suitable field re-definition,  which appeared to resurrect the validity
 of the analysis made in \cite{Hyunetal}. Unfortunately, the problem 
 still persists: By such a field  re-definition the higher derivatives 
 are simply shifted into the SUSY transformation laws and one must 
 reanalyze the issue with such modifications. 

Thus one sees that although the existing analyses are highly plausible they 
 are not air-tight. In view of the importance of precise understanding 
 of the role of supersymmetry and its connection with gauge symmetry,
 it is desirable to perform an unambiguous off-shell analysis with 
 arbitrary backgrounds. This motivates us to the study of the Ward identity,
to be described in the next two sections. 
\section{SUSY Ward Identity for the Effective Action in the 
Background Gauge}
Having argued the importance of off-shell analysis for arbitrary 
 trajectories, we shall now derive the SUSY Ward identity 
 for the effective action $\Gatil$  in the standard  background gauge, 
(\ref{sbgg}), used exclusively in the actual computations. 

To make use of the well-established method, let us 
 further split the quantum fluctuations $Y_m$ and $\Psi_\al$ into 
two parts, the diagonal and the off-diagonal, in the manner 
\begin{eqnarray}
Y_{m,ij} = {\yhat_{m,i}\over g}\delta_{ij} + \Yhat_{m,ij}\comma \qquad 
\Psi_{\al,ij} = {\psihat_{\al,i}\over g}\delta_{ij} + \Psihat_{\al,ij}\comma
\end{eqnarray}
and introduce the sources only for the diagonal fields:
\begin{eqnarray}
\Stil_s = \int\! d\tau \left( J_{m,i} \yhat_{m,i} 
+ \eta_{\al,i} \psihat_{\al,i}\right)
\period
\end{eqnarray}
The Euclidean generating functionals are defined by 
\begin{eqnarray}
\Ztil[J,\eta] &=& \int\! \calD \mu \, 
\exp\left( -\Stil_{tot}\right)
= \exp\left( -\Wtil[J,\eta]\right) \comma \\
\Stil_{tot} &\equiv & \Stil_0+\Stil_{gg} + \Stil_s \comma \qquad 
\calD \mu \equiv  \calD\!\Atil\, \calD Y\, \calD \Psi\, \calD C\, 
\calD \Cbar \comma  \nn
\end{eqnarray}
where $\Wtil[J,\eta]$ is the one for the connected functions. 
By making the change of integration variables corresponding 
to the supersymmetry  transformations, one obtains the primitive form
 of the Ward identity
\begin{eqnarray}
0=\langle \delta_\ep \Stil_{gg}\rangle + \langle \delta_\ep \Stil_s\rangle 
\period \label{primitive}
\end{eqnarray}
Here and in what follows, 
$\langle \calO \rangle$ for an operator $\calO$ means 
\begin{eqnarray}
\langle \calO \rangle = {
\displaystyle\int\! \calD \mu\, \calO 
e^{-\Stil_{tot}} \over \displaystyle\int\! \calD \mu\,
e^{-\Stil_{tot}} }  \period \label{expo}
\end{eqnarray}

We now rewrite this identity (\ref{primitive}) in terms of the  generating 
 functional $\Gatil$, which is 1PI (1-particle-irreducible) with respect 
 to the diagonal fields. Define as usual the classical fields and 
 $\Gatil$ by 
\begin{eqnarray}
y_{m,i(\tau)} &\equiv & {\delta \Wtil \over \delta J_{m,i}(\tau)}\comma 
\qquad \psi_{\al,i}(\tau) \equiv {\delta \Wtil \over \delta \eta_{\al,i}(\tau)}
\comma \\
\Gatil[y,\psi] &\equiv & W[J,\eta] -\int\! d\tau (J_{m,i}y_{m,i} 
+\eta_{\al,i}\psi_{\al,i})\period
\end{eqnarray}
Then the sources are expressed in terms of $\Gatil$ as
\begin{eqnarray}
J_{m,i}(\tau) = -{\delta \Gatil \over \delta y_{m,i}(\tau)}\comma \qquad 
 \eta_{\al,i}(\tau) = {\delta \Gatil \over \delta \psi_{\al,i}(\tau)}
\period
\end{eqnarray}
Therefore, the contribution to the Ward identity from the variation of the 
 source action can be written as
\begin{eqnarray}
\langle \delta_\ep \Stil_s \rangle = \int\! d\tau \left( 
 -{\delta \Gatil \over \delta y_{m,i}(\tau)} \langle \delta_\ep 
\yhat_{m,i}(\tau)\rangle 
+ {\delta \Gatil \over \delta \psi_{\al,i}(\tau)} \langle \delta_\ep 
 \psihat_{\al,i}(\tau) \rangle\right) \period \label{contsc}
\end{eqnarray}

As for the contribution $\langle \delta_\ep \Stil_{gg}\rangle$
 from the gauge-ghost part, a direct calculation yields a rather complicated 
 expression, which constitutes an inhomogeneous term in the Ward 
 identity regarded as a functional integro-differential equation for $\Gatil$. 
 This is undesirable since what we wish to understand 
 is how the supersymmetry acts on 
 the effective action $\Gatil$. 
Fortunately, it was noted long ago 
\cite{DF} in the context of four-dimensional super Yang-Mills theory 
 that one can reexpress such a term
 in a form similar to (\ref{contsc}). 
The first step is to note that the supersymmetry transformations (\ref{susyb})
 and (\ref{susyf}) commute with the BRST transformation (\ref{brstr}) on 
 all the fields,  as can be checked straightforwardly. Therefore, 
starting from (\ref{sgg}), 
$\langle \delta_\ep \Stil_{gg}\rangle$ can be written as 
\begin{eqnarray}
\langle \delta_\ep \Stil_{gg}\rangle &=& \langle \delta_\ep 
\delta_B \trace \int\! d\tau \left[ {1\over i} \Cbar \left(\Gtil-
{1 \over 2} b\right) \right] \rangle \nn\\
&=& \langle  
\delta_B \delta_\ep \trace \int\! d\tau \left[ {1\over i} \Cbar \left(\Gtil-
{1 \over 2} b\right) \right] \rangle 
= \langle \delta_B \calO_\ep\rangle \comma 
\end{eqnarray}
where
\begin{eqnarray}
\calO_\ep \equiv  {1\over i}\trace \int\! d\tau \Cbar \delta_\ep \Gtil
\end{eqnarray}
is a fermionic composite operator. 
This expression, being an expectation value of a BRST-exact form, vanishes 
 in the ordinary vacuum. However, in the presence of external sources, 
 it becomes proportional to the sources, and hence to the functional 
derivatives of $\Gatil$. Let us  collectively denote by $\phi$ and $J$ 
the basic fields and the corresponding sources respectively and 
 consider the generating functional with a source $j$ for the operator
 $\calO_\ep$:
\begin{eqnarray}
Z[J,j] =\int\! \calD\phi\, e^{-(S_{tot}+J\phi +j\calO_\ep)}\period
\end{eqnarray}
Now make a change of variables corresponding to the BRST transformation.
 We get
\begin{eqnarray}
0 =\int\! \calD \phi \, ((-1)^{|\phi|} J\delta_B\phi 
 - j\delta_B \calO_\ep) e^{-(S_{tot}+J\phi +j\calO_\ep)} \comma 
\end{eqnarray}
where $|\phi|$ is 0 (1) if $\phi$ is bosonic (fermionic). 
By differentiating with respect to $j$ once, setting $j=0$, and then 
expressing the source $J$ in terms of $\Gatil$, 
 one easily obtains the following  BRST Ward identity:
\begin{eqnarray}
\langle \delta_B \calO_\ep\rangle &=& - \int\! d\tau {\delta \Gatil \over \delta
\phi(\tau)} \langle \delta_B \phi(\tau) \calO_\ep \rangle \period
\end{eqnarray}
In this way, we get 
\begin{eqnarray}
\langle \delta_\ep \Stil_{gg} \rangle 
 = -\int\! d\tau \left( {\delta \Gatil \over \delta y_{m,i}(\tau)}
 \langle \delta_B \yhat_{m,i}(\tau) \calO_\ep\rangle 
 + {\delta \Gatil \over \delta \psi_{\al,i}(\tau)}
 \langle \delta_B \psihat_{\al,i}(\tau) \calO_\ep\rangle \right)
\period
\end{eqnarray}

Putting all together, we arrive at the following SUSY Ward identity 
 expressed solely in terms of the derivatives of $\Gatil$:
\begin{eqnarray}
0&=& \int\! d\tau \Biggl( {\delta \Gatil \over \delta y_{m,i}(\tau)}
(\langle \delta_\ep \yhat_{m,i}(\tau) \rangle +
 \langle \delta_B \yhat_{m,i}(\tau) \calO_\ep\rangle )\nn\\
&& \qquad  -{\delta \Gatil \over \delta \psi_{\al,i}(\tau)}
 (\langle \delta_\ep \psihat_{\al,i}(\tau)\rangle
-\langle \delta_B \psihat_{\al,i}(\tau) \calO_\ep\rangle)\Biggr)\period
\label{wardone}
\end{eqnarray}

Normally it is now a simple matter to convert this into the desired 
Ward identity for the effective action as 
 a functional of the backgrounds $B^m$ and $\theta_\al$: One would 
rewrite the derivatives with respect to $y_m$ and $\psi_\al$ into 
those with respect to $B^m$ and $\theta_\al$ and then set 
 $y_m =\psi_\al=0$. This  procedure is indeed valid for the fermions
 since $\psihat_\al$ and $\theta_\al$ always appear in the combination 
 $\psihat_\al +\theta_\al$ in the original action 
 and it is a simple matter to prove that this gets converted to 
 $\psi_\al + \theta_\al$ in $\Gatil$.

 This is {\it not} so for the bosonic field. 
While most of the $B^m$ dependence comes from the splitting 
 $X_m = (1/g) (B_m +\yhat_m) + \Yhat_m$, where $B_m$ and $\yhat_m$ appear 
 together, $B^m$ in the gauge-ghost sector (in the standard background 
 gauge) is not accompanied with $\yhat_m$. From the point of view of the 
 Ward identity above, it is an independent extra parameter field which 
 should be distinguished from the bona fide  background field. 
This is why we chose to start out with a different symbol $\Btil^m$ 
for this field. 

Now the problem we face is the following. In order to be able to 
 apply  the Ward identity to the case of the standard background gauge,
 we need to 
express the dependence on $\Btil^m$ again in the form of the functional
 derivative of $\Gatil$ with respect to $B^m$. In other words, we must 
 disentangle the two different types of $B^m$ dependence buried in the 
 standard background gauge formulation and construct the Ward identity which
 takes both types of dependence into account. Although this is a 
technical rather than a conceptual problem, it is again a manifestation
 of the gauge theory nature of the Matrix theory, which has often been 
neglected. 

The problem  can be solved as follows. Since $\Btil^m$ appears only 
 in the gauge-ghost action and the variation with respect to it commutes 
 with the BRST transformation, we get from (\ref{sgg}) 
\begin{eqnarray}
{\delta \Stil_{gg} \over \delta \Btil^{m,i}(\tau)} = \delta_B 
 \calO_{m,i}(\tau) \comma 
\end{eqnarray}
where
\begin{eqnarray}
 \calO_{m,i} = \sum_j\left( \Cbar_{ji}Y_{m,ij}-\Cbar_{ij} Y_{m,ji}
\right) \comma 
\end{eqnarray}
and $\Btil^{m,i}$ stands for the diagonal elements $\Btil^m_{ii}$. 
The expectation value of the left-hand side
can be expressed in terms of the generating functional $\Wtil$ as 
$\delta \Wtil/\delta \Btil_{m,i}$, which is equal to $\delta \Gatil/
\delta \Btil_{m,i}$ since it is a parameter field. On the other hand
 the expectation value of the right-hand side can be treated in exactly the
 same way as we treated $\delta_B \calO_\ep$. In this way  we get 
 the identity
\begin{eqnarray}
{\delta \Gatil \over \delta \Btil_{m,i}(\tau)} 
 &=& -\int\! d\tau' \Biggl( {\delta \Gatil \over \delta y_{n,j}(\tau')}
 \langle \delta_B \yhat_{n,j}(\tau') \calO_{m,i}(\tau)\rangle \nn\\
&& \qquad  + {\delta \Gatil \over \delta \psi_{\al,j}(\tau')}
 \langle \delta_B \psihat_{\al,j}(\tau') \calO_{m,i}(\tau)\rangle \Biggr)
\period \label{delbtil}
\end{eqnarray}

Now let us replace $y_m$ and $\psi_\al$ with $B_m$ and $\theta_\al$ 
 respectively and then set $y_m=\psi_\al=0$ in the previous Ward identity
 ({\ref{wardone}) and in the relation (\ref{delbtil}) above. 
In the limit $\Btil_m \rightarrow B_m$, the {\it total} variation of $\Gatil$
 with respect to $B_m$, which we denote by $\Delta \Gatil/\Delta B_m$ 
 to avoid confusion, is 
\begin{eqnarray}
{\Delta \Gatil\over \Delta B_{m,i}(\tau)} = \left(
 {\delta \Gatil \over \delta B_{m,i}(\tau)} + 
{\delta \Gatil \over \delta \Btil_{m,i}(\tau)} \right)_{\Btil =B}
\period
\end{eqnarray}
Substituting  (\ref{delbtil}) we then get 
\begin{eqnarray}
{\Delta \Gatil\over \Delta B_{m,i}(\tau)} &=&
\int\! d\tau' {\delta \Gatil \over \delta B_{n,j}(\tau') }
 T _{nj;mi}(\tau,\tau')\nn\\
&& \qquad  -\int\! d\tau' {\delta \Gatil \over \delta \theta_{\al,j}(\tau')}
 \langle \delta_B \psihat_{\al,j}(\tau') \calO_{m,i}(\tau)\rangle \comma 
\end{eqnarray}
with 
\begin{eqnarray}
 T _{nj;mi}(\tau,\tau') &\equiv & 
\delta_{mn}\delta_{ij}\delta (\tau-\tau') 
 -\langle \delta_B \yhat_{n,j}(\tau') \calO_{m,i}(\tau)\rangle \period
\end{eqnarray}
By inverting this relation, we can express the partial variation
 $\delta\Gatil/\delta B_m$ in terms of the total variation 
$\Delta \Gatil/\Delta B_m$:
\begin{eqnarray}
{\delta \Gatil \over \delta B_{m,i}(\tau)} 
 &=& \int\! d\tau'  T ^{-1}_{mi,nj}(\tau,\tau') {\Delta \Gatil 
\over \Delta B_{n,j}(\tau')} \nn\\
&& \qquad + \int\! d\tau'  T ^{-1}_{mi,nj}(\tau,\tau')
\int\! d\tau'' {\delta \Gatil \over \delta \theta_{\al,k}(\tau'')}
 \langle \delta_B \psi_{\al,k}(\tau'') \calO_{n,j}(\tau')\rangle
\comma
\end{eqnarray}
where $ T ^{-1}$ is the inverse of $ T $. 

Finally, the correct Ward identity in the standard background gauge 
 is obtained by substituting this expression into (\ref{wardone}).
Once the limit $\Btil_m \rightarrow B_m$ is taken, the total variation 
 $\Delta \Gatil/\Delta B_m$ can be identified with 
$\delta \Gatil/\delta B_m$, which denotes 
the usual functional derivative of the effective action computed 
 in the standard background gauge (\ie with $B_m$ in the gauge-fixing 
 term.) With this understood, the result can be put in the desired form
\begin{eqnarray}
0 = \int\! d\tau 
\left( \Delta_\ep B_{m,i}(\tau){\delta \Gatil \over \delta B_{n,j}(\tau)}
 + \Delta_\ep \theta_{\al,i}(\tau)
{\delta\Gatil \over \delta \theta_{\al,i}(\tau) }  \right)\comma 
\end{eqnarray}
where the effective SUSY transformation laws are given by 
\begin{eqnarray}
\Delta_\ep B_{m,i}(\tau) &=& \int\! d\tau'  T ^{-1}_{mi,nj}(\tau',\tau)
(\langle \delta_\ep \yhat_{n,j}(\tau') \rangle +
 \langle \delta_B \yhat_{n,j}(\tau') \calO_\ep\rangle )\comma 
\label{delb} \\
\Delta_\ep \theta_{\al,i}(\tau) &=&
 \langle \delta_\ep \psihat_{\al,i}(\tau)\rangle 
 -\langle \delta_B \psihat_{\al,i}(\tau)\calO_\ep\rangle \nn\\
&& - \int\! d\tau' d\tau''  T ^{-1}_{mk,nj}(\tau'',\tau') 
\langle \delta_B \psihat_{\al,i}(\tau) \calO_{n,j}(\tau') \rangle
 \langle \delta_\ep \yhat_{m,k}(\tau'') 
+ \delta_B \yhat_{m,k}(\tau'')\calO_\ep
\rangle \period \nn\\ \label{delth}
\end{eqnarray}
(As said before $y=\psi=0$ is understood.)
This is the main result of this section. 

Note the following features:
\begin{itemize}
	\item We have succeeded in putting the Ward identity in the 
 form where the supersymmetry transformation laws for the effective 
 action are cleanly identified in closed forms. 
	\item As expected, the supersymmetry and the gauge (BRST) symmetry
 are non-trivially intertwined. Naively, one might expect that the 
 effective transformation laws are obtained as the expectation values 
 of the original transformation laws (\ref{susyb}) and (\ref{susyf}). 
In our notation, they are represented by
 $\langle \delta_\ep \yhat_{n,j}\rangle$ and 
$\langle \delta_\ep\psihat_{\al,i}\rangle$ in $\Delta_\ep B_{m,i}$ and 
 $\Delta_\ep \theta_{\al,i}$ respectively. The actual transformation 
 laws, (\ref{delb}) and (\ref{delth}), are much more complicated. One can 
 see, however, that the corrections to the naive laws all involve 
 $\delta_B$, \ie the BRST transformation. Since the quantization of the 
 system inevitably requires a gauge fixing, this is a {\it universal} feature, 
 not special to the standard background gauge adopted here. 
	\item As has already been remarked, the transformation laws derived above
 are {\it exact}, albeit somewhat formal at this stage.
 In particular, there is no inherent distinction between the tree 
level contribution and the quantum corrections. Thus it is far from 
 obvious that the anti-commutator of the effective transformations
 would close solely on the translation generator, as it does at the tree 
 level: We have so far not been able to produce a proof. 
\end{itemize}

In the next section, we will carefully examine the structure of 
this Ward identity at the 1-loop level and draw implications. 
\section{Explicit Calculations and Implications}
We now compute the effective SUSY transformation laws explicitly and 
 study the implications of the Ward identity. 
\subsection{Source-Probe Situation}
Although the Ward identity derived in the previous section is valid 
 for any background, we shall restrict ourselves to
 the source-probe situation, since the existing calculations of
 the effective action itself, to be compared later, are more complete for 
 this configuration.  The background fields representing this situation
 for the gauge group $U(N+1)$ were  already described in (\ref{sourceprobe}), 
 which we display again for convenience:
\begin{eqnarray}
B_m = {\rm diag}\, (r_m, 0,0, \ldots , 0) \comma \qquad 
 \theta_\al = {\rm diag}\, (\theta_\al,0,0,\ldots, 0) \period
\end{eqnarray}
$r_m$ represents the coordinate of the probe and $\theta_\al$ its 
 spin content. We shall denote the time derivative
 of the coordinate by $v_m \equiv \del_\tau r_m$.

To facilitate the computations, it is convenient to introduce the 
 following notations:
For any  matrix $U$ we define $U_I \equiv U_{1I}\comma \
 U^\ast_I \equiv U_{I1}$, where $I=2\sim N+1$.  (The symbol ${}^\ast$ here 
 does not stand for complex conjugation.) We shall call such index $I$ 
 the off-diagonal matrix vector index. 
Then the \lq\lq 11" component 
 of a product of two matrices becomes $(UV)_{11} = 
 \sum_{I=2}^{N+1} U_I V^\ast_I$, which we abbreviate as $U\cdot V^\ast$. 

With this convention, the basic quantities appearing in the Ward 
 identity take the following forms:
\begin{eqnarray} 
\langle \delta_\ep \yhat_m\rangle &=& -i\ep^T \ga_m \theta \comma \\
 \langle \delta_\ep \psihat_\al \rangle
&=& i\left(\! \vslash _{\al\be} -ig^2 \ga^m_{\al\be} 
\langle A\cdot Y^\ast_m -
A^\ast \cdot Y_m\rangle + {g^2\over 2}\ga^{mn}_{\al\be}
\langle Y_m\cdot Y^\ast_n -Y_m^\ast\cdot  Y_n\rangle\!\right)\ep_\be \comma \\
\delta_B \yhat_m &=& -ig^2 (Y_m\cdot C^\ast -Y_m^\ast\cdot C) \comma \\
\delta_B \psihat_\al &=& ig^2(C\cdot \Psi^\ast_\al -C^\ast \cdot \Psi_\al)
\comma  \\
\calO_\ep &=& -i\ep_\al \int\! d\tau \Biggl(
{1\over g} \Cbar_{11}(\thdot_\al +\dot{\psihat \ }\mbox{}\!\!_\al) 
 + \Cbar \cdot (\del_\tau + \rslash)_{\al\be}\Psi^\ast_\be
 + \Cbar^\ast \cdot (\del_\tau -\rslash)_{\al\be}\Psi_\be\nn\\
&&\hspace{4cm} + \Cbar_{IJ} \dot{\Theta}_{JI} \Biggr) \comma \\
\calO_m &=& \Cbar^\ast \cdot Y_m-\Cbar \cdot Y_m^\ast  \period
\end{eqnarray}
%
\subsection{One-Loop at Order Two}
In this article, we shall perform the calculation of the simplest 
non-trivial contributions to the effective transformation laws,
 namely those which govern the order 2 part of the 1-loop effective 
action. Here the order is defined as in (\ref{deforder}), \ie 
the number of derivatives plus twice the number of fermions.
Since the tree action is already of 
 order 2, this means that we need to compute 
 $\Delta_\ep r_m$ and $\Delta_\ep \theta_\al$ to order 0 at 1-loop, 
 where the transformation parameters $\ep_\al$ are considered to be of 
 order $-1/2$. 

Such contributions are further classified according to 
 the number of $\theta$'s involved. 
Because $r_m$ and $\theta_\al$ 
 are of order 0 and $1/2$ respectively, we need to consider 
$\Delta_\ep r_m$ to linear order in $\theta$, 
while for $\Delta_\ep \theta_\al$ we need $\theta^2$ contributions as well.
 In what follows, it is more convenient to refer only to the number 
 of additional $\theta$'s relative to the tree level contribution. 
For instance, \lq\lq a correction to $\calO(\theta^0)$"  for $\Delta_\ep
r_m$ refers to a term of the form $\ep\theta f(r)$. 

At 1-loop order, since the terms involving the BRST variation $\delta_B$ 
 start only at 1-loop, the expressions (\ref{delb}) and 
 (\ref{delth}) for $\Delta_\ep r_m$ and  $\Delta_\ep \theta$ simplify to 
\begin{eqnarray}
\Delta_\ep r^m (\tau) &=& {1\over i} (\ep \ga^m \theta)(\tau) 
 +\int\!d\tau' \langle \delta_B\yhat_m(\tau) \calO_n(\tau')\rangle
 {1\over i}(\ep \ga^n\theta)(\tau') +
 \langle \delta_B \yhat_m(\tau) \calO_\ep \rangle\comma  \label{delb1}\\
\Delta_\ep\theta_\al(\tau) &=& \langle \delta_\ep \psihat_\al(\tau)\rangle
 -\langle\delta_B\psihat_\al(\tau) \calO_\ep\rangle 
 -\int\! d\tau' \langle \delta_B \psihat_\al(\tau) \calO_m(\tau')\rangle
{1\over i} (\ep \ga^m\theta)(\tau')\period \label{delth1}
\end{eqnarray}
The expectation values of the composite operators themselves
simplify considerably at this order. 
It is easy to check that at 1-loop only 
 the 2-point functions contribute. Moreover, since there are no 
 mixing between the fields with different off-diagonal matrix vector 
 indices in such propagators, we always have the structure $\langle U_I(\tau) 
 V^\ast_J(\tau') \rangle = \delta_{IJ} \langle U(\tau) V^\ast(\tau')\rangle$,
 where $U$ and $V^\ast$ can be regarded as  single-component fields. 
$\delta_{IJ}$'s then contract to produce a factor of $N$, which goes with 
 the coupling $g^2$. With these remarks, the relevant multi-body 
 expectation values become 
\begin{eqnarray}
\langle \delta_\ep \psihat_\al(\tau) \rangle
&=& i (\vslash \ep)_\al(\tau) +g^2N \Biggl(\ga^m_{\al\be} 
\langle A(\tau)Y^\ast_m(\tau) -
A^\ast(\tau) Y_m(\tau)\rangle \nn\\
&& + {i\over 2}\ga^{mn}_{\al\be}
\langle Y_m (\tau)Y^\ast_n(\tau) -Y_m^\ast(\tau) Y_n(\tau)
\rangle \Biggr)\ep_\be(\tau) \comma \nn\\
\\
\langle \delta_B \yhat_n(\tau) \calO_\ep\rangle 
 &=& \ep_\al g^2N\!\int\! d\tau' \Biggl( \langle C^\ast(\tau)\Cbar(\tau')
\rangle (\del +\rslash)_{\al\be} \langle \Psi^\ast_\be(\tau')
 Y_m(\tau)\rangle  \nn\\
&& -\langle C(\tau) \Cbar^\ast(\tau')\rangle (\del
 -\rslash)_{\al\be} \langle \Psi_\be(\tau') Y^\ast_m(\tau) 
\rangle\Biggr)\comma \label{byo}\\
\langle \delta_B\psihat_\be(\tau)\calO_\ep\rangle &=& \ep_\al g^2N
\!\int\! d\tau' \Biggl(  \langle C^\ast(\tau)\Cbar(\tau')
\rangle(\del +\rslash)_{\al\ga} \langle \Psi^\ast_\ga (\tau')  
 \Psi_\be(\tau)\rangle \nn\\
&& -\langle C(\tau) \Cbar^\ast(\tau')\rangle
(\del -\rslash)_{\al\ga}\langle \Psi_\ga(\tau')
 \Psi^\ast_\be(\tau)\rangle \Biggr)\nn\\
&& -g \ep_\be\!\int\! d\tau^\prime
\langle C\Psi^\ast(\tau)\Cbar_{11}
\dot{\widehat{\psi}\ }\mbox{}\!\!_\beta(\tau^\prime)
\rangle  
+ g \ep_\be\!\int\! d\tau^\prime
\langle C^\ast\Psi(\tau)\Cbar_{11}
\dot{\widehat{\psi}\ }\mbox{}\!\!_\beta(\tau^\prime)\rangle \comma \nn \\
\\ 
\langle \delta_B \yhat_n(\tau') \calO_m(\tau)\rangle 
&=& -ig^2N \Biggl( \langle Y_m (\tau') Y^\ast_n(\tau)\rangle 
\langle(C^\ast(\tau') \Cbar (\tau)\rangle \nn\\
&& + \langle Y^\ast_m (\tau') Y_n(\tau)\rangle 
\langle(C(\tau') \Cbar^\ast (\tau)\rangle \Biggr)\comma 
\end{eqnarray}
where $\del$ stands for the derivative with respect to $\tau$. We will 
 now evaluate these expressions to the relevant order by perturbation 
 theory. 
\subsubsection{Corrections at $\calO(\theta^0)$}
Let us begin with the corrections at $\calO(\theta^0)$. 
First we need to compute the tree-level propagators. 
The part of the Lagrangian quadratic in the fields without $\theta$ is 
 of the form 
\begin{eqnarray}
\calL &=& Y_m\cdot (-\del^2 +r^2) Y_m^\ast + \Atil \cdot(-\del^2+r^2)\Atil^\ast
 -2iv_m(\Atil\cdot Y_m^\ast -\Atil^\ast \cdot Y_m) \nn\\
&& +{1\over 2g^2} \psihat \del \psihat 
+\Psi_\al \cdot (\del+\rslash)_{\al\be} \Psi^\ast_\be \nn\\
&& + i\Cbar\cdot(-\del^2+r^2)C^\ast +i\Cbar^\ast\cdot (-\del^2+r^2)C
+i \Cbar_{11} (-\del^2) C_{11}
\period
\end{eqnarray}
Following the remark already made on the trivial dependence on the 
off-diagonal matrix vector indices, we may treat the fields $Y_m, 
 \Atil, \Psi_\al$ and $C, \Cbar$  in $\calL$ 
 as if they were single-component. The propagators for the massless fields
 $\psihat_\al$ and $C_{11}, \Cbar_{11}$ turn out to be needed only 
 at  $\calO(\theta^2)$, to be discussed in the next subsection.

The simplest is the ghost propagator, which can be directly  read off as 
\begin{eqnarray}
&& \langle \Cbar(\tau)C^\ast(\tau')\rangle =
\langle \Cbar^\ast(\tau) C(\tau')\rangle \nn\\
&& \qquad = -\langle C^\ast(\tau) \Cbar(\tau')
\rangle 
 =-\langle C(\tau) \Cbar^\ast(\tau')\rangle = i\bra{\tau} \Delta
 \ket{\tau'} \comma 
\end{eqnarray}
where
\begin{eqnarray}
\Delta &\equiv & (-\del^2+r^2)^{-1}\period
\end{eqnarray}

The propagators for the $\Psi_\al$ system are given by 
\begin{eqnarray}
\langle \Psi_\al(\tau)\Psi^\ast_\be(\tau')\rangle &=&\bra{\tau}
 D^-_{\al\be} \ket{\tau'} \comma \label{PsiPsiast}\\
\langle \Psi_\al^\ast(\tau)\Psi_\be(\tau')\rangle &=&\bra{\tau}
 D^+_{\al\be} \ket{\tau'} \comma 
\end{eqnarray}
where
\begin{eqnarray}
D^\pm &\equiv & (\del \pm \rslash)^{-1} \period
\end{eqnarray}
>From the relation $-(\del \pm \rslash)(\del \mp \rslash) 
=(-\del^2+r^2)(1\pm \Delta \vslash)$, we can expand $D^\pm$ in powers of
 the velocity $v(\tau)$ as 
\begin{eqnarray}
D^\pm &=& -(\del \mp \rslash) (1 \mp \Delta \vslash)^{-1} \Delta
= -(\del \mp \rslash)(\Delta \pm \Delta \vslash \Delta + \cdots)
\period
\end{eqnarray}
To the order of interest, we will only need the $v$-independent part. 

Now as for the $Y_m$-$\Atil$ system, we have a mixing term with an arbitrary 
 coefficient function $v_m(\tau)$ and it can only be resolved in the 
 derivative expansion. Expanding in powers of the mixing term, we readily
 obtain 
\begin{eqnarray}
\langle Y^m(\tau)^\ast Y^n(\tau')\rangle & =&\delta_{mn} \bra{\tau} \Delta 
  \ket{\tau'} + \calO(v^2)\comma \nn\\
\langle \Atil(\tau) \Atil^\ast(\tau')\rangle &=& 
\bra{\tau} \Delta  \ket{\tau'} + \calO(v^2)\comma 
\label{AAast}\\
\langle Y_m(\tau)\Atil^\ast(\tau')\rangle &=& -\langle \Atil(\tau) 
 Y_m^\ast (\tau')\rangle =\bra{\tau} 2i\Delta v_m \Delta 
  \ket{\tau'} + \calO(v^3)\period\nn
\end{eqnarray}

The only other 2-point functions appearing at this order are 
$\langle \Psi_\al(\tau) Y_m^\ast(\tau')\rangle $ and \\
$\langle \Psi^\ast_{\al}(\tau)Y_m (\tau')\rangle $, which can be readily 
 computed by inserting the vertices of the form 
\begin{eqnarray}
-\theta_\al\ga^m_{\al\be} (Y_m \cdot\Psi^\ast_\be -Y^\ast_m\cdot \Psi_\be)
 \period 
\end{eqnarray}
The results are
\begin{eqnarray}
\langle \Psi_\al(\tau) Y_m^\ast(\tau')\rangle 
 &=& -\bra{\tau} D^-_{\al\be} (\ga^m\theta)_\be \Delta \ket{\tau'} \comma \\
\langle \Psi^\ast_{\al}(\tau)Y_m (\tau')\rangle 
&=& \bra{\tau} D^+_{\al\be} (\ga^m\theta)_\be \Delta \ket{\tau'} \period
\end{eqnarray}

With this preparation, the calculations of the various expectation values
 in the Ward identity can be performed efficiently to the desired 
 order with the use of the so-called \lq\lq normal ordering method"
 developed in \cite{okawa9903}. 
 The essence of this method is to first rearrange 
 the order of a product of various operators and functions into 
  the standard form $\sim f(\tau) \del^m \Delta^n$,
 using the commutation relations 
 such as $\com{\Delta}{\del} = 2\Delta (r\cdot v) \Delta$ etc., and 
 then use Baker-Campbell-Hausdorff and the Gaussian 
 integration formulas to evaluate it. A useful list of formulas so 
 obtained are collected in the appendix of \cite{hm9904}.
Below, we shall give a sample calculation of this sort and then simply 
 list the results for the needed expectation values. 

As an example, let us consider
 $\langle \delta_B \yhat_n(\tau) \calO_\ep\rangle$ 
 given in (\ref{byo}). Substituting the expressions for the various 
 propagators already computed, it becomes
\begin{eqnarray}
\langle \delta_B \yhat^n(\tau) \calO_\ep\rangle
&=& -i\ep^T g^2N \bra{\tau} \Delta [(\del+\rslash) D^+
+(\del-\rslash)D^-]\ga^n\theta \Delta \ket{\tau} \nn\\
&=& -2i\ep^T g^2N \bra{\tau} \Delta (-\del^2+r^2)\Delta
\ga^n\theta \Delta \ket{\tau} \period
\end{eqnarray}
To the order of interest, the normal-ordering is trivial and we get
\begin{eqnarray}
\langle \delta_B \yhat_n(\tau) \calO_\ep\rangle
= -2i\ep^T g^2N \ga_n\theta \bra{\tau} \Delta^2\ket{\tau} =-i \ep^T 
\ga_n\theta \half {g^2N\over r^3} \period
\end{eqnarray}
In a similar manner, we obtain the following results:
\begin{eqnarray}
\langle \delta_B \yhat_n(\tau') \calO_m(\tau)\rangle &=& -2g^2N\delta_{nm}
 \bra{\tau'} \Delta \ket{\tau} ^2 \comma \\
\int\! d\tau' \langle \delta_B \yhat_m(\tau) \calO_n(\tau')
{1\over i}(\ep\ga^n\theta)(\tau') 
 &=& -i \ep \ga_m\theta \half {g^2N\over r^3} \comma \\
\langle \delta_B \psihat_\be(\tau) \calO_\ep\rangle &=& 0 \comma \\
\langle \delta_B \yhat_m(\tau)\calO_\ep\rangle &=&
 -i \ep \ga_m\theta \half {g^2N\over r^3} \comma \\
\langle A\cdot Y_m^\ast -A^\ast \cdot Y_m\rangle &=& 
{N\over i}{v_m \over r^3} \comma \\
\langle Y_m\cdot Y_n^\ast -Y_m^\ast\cdot Y_n \rangle &=& 0 \period
\end{eqnarray}
Making use of these formulas, we find the effective SUSY transformation 
laws at $\calO(\theta^0)$ to be 
\begin{eqnarray}
\Delta_\ep r^m &=& -i\ep^T \ga^m\theta \left( 1+ {g^2N \over r^3}\right)\comma 
 \label{deltar}\\
\Delta_\ep \theta &=& i \vslash  \ep
\left( 1-{g^2N\over r^3}\right)\period \label{deltath}
\end{eqnarray}
Already at this order, there are non-trivial corrections to 
 the tree level laws. 
For $\Delta_\ep \theta$ the entire correction 
 came from the expectation value of the non-linear part of the original 
 SUSY transformation law, namely from $-ig  \com{\Atil}{Y_m} \ga^m$
 contained in  $\com{ D_\tau}{X_m} \ga^m$. On the other hand, the 
 same procedure is not applicable for $\Delta_\ep r_m$: 
If one naively 
 took the expectation value of the basic transformation law $\delta_\ep X^m 
 = -i\ep\ga^m \Theta$, one would not get any corrections to all orders. 
What we want is the effective 
 SUSY transformation {\it operating on the effective action $\Gatil$}
 and this can only be obtained through the analysis of the Ward identity, 
 as we have done.  One can see 
 from the calculations outlined above that 
 exactly half of the quantum correction 
 for $\Delta_\ep r^m$ came from $\langle \delta_\ep \Stil_{gg}\rangle$ and 
the other half was produced from the procedure of taking into account 
the dependence on the extra parameter field we originally called $\Btil_m$. 
\subsection{Corrections at $\calO(\theta^2)$}
Now we move on to the corrections at $\calO(\theta^2)$. Since $\theta^2$ is
 of order 1, we  need to compute such contributions only for 
$\Delta_\ep \theta_\al$. The procedure is entirely similar to 
 the $\calO(\theta^0)$ case but the calculations are more involved and 
 we relegate the details to the Appendix A. As shown there, 
 7 types of diagrams contribute. Two of them, 
 diagrams (B2-a,b),  involve genuine 3 point vertices as well as massless 
 propagators given by 
\begin{eqnarray}
\langle \psihat_\al(\tau) \psihat_\be(\tau')\rangle = g^2 \langle \tau |
{1\over \del} | \tau'\rangle \comma \qquad 
\langle C_{11}(\tau) \Cbar_{11}(\tau')\rangle = i\langle \tau |
 {1\over \del^2} | \tau'\rangle\period
\end{eqnarray}
They are singular in the infrared, but such singularities cancel in 
 the end result. After some calculations, we find the $\calO(\theta^2)$ 
 corrections to $\Delta_\ep \theta_\al$ to be 
\begin{eqnarray}
\Delta^{\theta^2}_\ep \theta_\al &=& {3ig^2N \over 16r^5}
\Biggl(-r_l (\theta\ga^{mnl}\theta) (\ga^{nm}\ep)_\al
 + 2r_l(\theta \ga^{ml}\theta) (\ga^m\ep)_\al -4r_l(\ep\ga^m\theta)
(\theta \ga^{ml})_\al \nn\\
&& \qquad +4(\ep\theta)(\rslash\theta)_\al 
 -4(\ep \rslash \theta)\theta_\al\Biggr) \period \label{th2}
\end{eqnarray}
Although it appears quite complicated, it can be drastically simplified 
with the use of the $SO(9)$ Fierz identities\cite{TR} described in 
 the Appendix B.  The relevant identity is 
\begin{eqnarray}
0&=&  -r_l(\theta \ga^{mnl}\theta)(\ga^{nm}\ep)_\al
 +2 r_l(\theta \ga^{nl}\theta)( \ga^{n}\ep)_\al
+4  r_l(\ep \ga^n \theta)(\ga^{nl}\theta)_\al \nn\\
&& \qquad +4r_l(\ep\theta)(\ga^l\theta)_\al + 12 r_l(\ep \ga^{l}\theta)
\theta_\al \period
\end{eqnarray}
Applying this to (\ref{th2}), we get
\begin{eqnarray}
\Delta^{\theta^2}_\ep \theta_\al = -{3ig^2N \over r^5}
(\ep \rslash \theta) \theta_\al \period
\end{eqnarray}
\parsmallskip
Let us summarize the results. To order 2 at 1-loop, 
 the effective supersymmetry transformations laws are found to be 
\begin{eqnarray}
\Delta_\ep r^m &=& -i\ep^T \ga^m\theta \left( 1+ {g^2N \over r^3}\right)\comma 
 \label{delrfin}\\
\Delta_\ep \theta_\al &=& i (\vslash  \ep)_\al
\left( 1-{g^2N\over r^3}\right)
 -{3ig^2N \over r^5}
(\ep \rslash \theta) \theta_\al \period\label{delthfin}
\end{eqnarray}
\subsubsection{Closure of the Effective Algebra and Field Redefinition}
It is of interest to examine the closure property of the transformation 
 laws obtained above. By a simple calculation, one immediately finds

\begin{eqnarray}
\com{\Delta_{\ep_1}}{\Delta_{\ep_2}} r_m &=& 2\dot{r}_m (\ep_2 \ep_1)
+\calO(g^4) \comma \\
\com{\Delta_{\ep_1}}{\Delta_{\ep_2}} \theta_\al &=& 2\thdot_\al (\ep_2 \ep_1)
+\calO(g^4) \period
\end{eqnarray}
Thus to the 1-loop order, the closure turned out to be precisely canonical. 

As we  
 remarked at the end of Sec.4, this is not a feature guaranteed by the general 
 analysis. One way to appreciate this is to consider a field redefinition
 which makes the form of $\Delta_\ep r^m$ to be the same as the one 
 at the tree level. 
On general grounds, the most general form of $\Delta_\ep r^m$ and 
 $\Delta_\ep \theta_\al$, 
 to the order we are considering, are
\begin{eqnarray}
\Delta_\ep r^m &=& {1\over i}(\ep \ga^m \theta) \left( 1+ g^2 F(r)\right)
\comma \label{delrgen}\\
\Delta_\ep \theta_\al &=& i(\vslash \ep)_\al + g^2 G_{\al\be}(r,\theta)
\ep_\be \period
\end{eqnarray}
>From (\ref{delrgen}), the desired field redefinition can be read off as
\begin{eqnarray}
\tilde{\theta}_\al \equiv \left( 1+ g^2 F(r)\right) \theta_\al \period
\end{eqnarray}
The transformation law for this new field is then
\begin{eqnarray}
\Delta_\ep \tilde{\theta}_\al = i(\vslash \ep)_\al 
 + g^2 \left( i(\vslash\ep)_\al F(r) +\Delta_\ep F(r)\tilde{\theta}_\al +
 G_{\al\be}(r, \tilde{\theta})\ep_\be
\right) + \calO(g^4) \period
\end{eqnarray}
It is quite non-trivial that the $\calO(g^2)$ part of this expression 
 vanishes exactly. 
\subsection{Implication of the Ward Identity}
Having found the transformation laws, we are now ready to 
 analyze the consequence of the Ward identity on the structure of $\Gatil$.

Let us write the effective action up to 1-loop as $\Gatil = \Gatil_0 
 + \Gatil_1$, the subscript denoting the number of loops. 
The tree-level action is given by 
\begin{eqnarray}
\Gatil_0 =\int\! d\tau \left( {v^2 \over 2g^2} + {\theta\thdot \over 2g^2}
\right) \period \label{treeaction}
\end{eqnarray}
As for $\Gatil_1$, it is easy to convince oneself that the most 
general structure at order 2, up to  total derivatives, is 
\begin{eqnarray}
\Gatil_1 &=& \int\! d\tau \Biggl( A{v^2\over r^3} + B{(v\cdot r)^2\over r^5}
 + C{\theta^T\thdot \over r^3} + D{\theta^T \rslash \thdot \over r^4}\nn\\
&&  + E{\theta^T \ga^{mn} r_mv_n \theta \over r^5}
+F {r^l r^k (\theta \ga^{lm}\theta)(\theta \ga^{mk}\theta)
 \over r^7}\Biggr) \comma \label{genonelp}
\end{eqnarray}
where $A\sim F$ are numerical constants\footnote{The structure 
 of the form $\sim r^l r^k (\theta \ga^{lmn}\theta)(\theta \ga^{mkn}\theta)$
 can be expressed in terms of the last term in (\ref{genonelp})
via a Fierz identity and 
 hence can be omitted.}. 

Now we demand that $\Delta_\ep \Gatil$ vanish to the order of interest. 
The simplest way to proceed is as follows. First, look at the $\calO(\theta^5)$
 terms, which can only be produced from the last term of 
 (\ref{genonelp}) by the tree-level part of the transformation 
$\Delta_\ep r_m$. They read
\begin{eqnarray}
\int\! d\tau\,  F \frac{i}{r^9}r^k \bigg(
-2r^2
(\epsilon\gamma^l\theta)
(\theta\gamma^{l m}\theta)
+7
r^l r^j (\epsilon\gamma^j\theta)
(\theta\gamma^{l m}\theta)
\bigg) (\theta\gamma^{mk}\theta)\period
\end{eqnarray}
It can be checked that the integrand does not vanish by any of the Fierz
 identities\footnote{We have also checked this numerically.}.
Thus we find $F=0$. Next, demand that the $\calO(\theta)$ part of $\Delta_\ep
 \Gatil$ vanish. It is straightforward to show that this reduces the 
allowed form of  $\Gatil_1$ to be 
\begin{eqnarray}
\Gatil_1 =\int\! d\tau \left(  A{v^2\over r^3} 
+(A+N){\theta\thdot \over r^3}
 +{3A\over 2}{\theta \ga^{mn} r_m v_n \theta \over r^5}\right) \comma 
\end{eqnarray}
where $A$ remains undetermined.  Finally, look at the $\calO(\theta^3)$ 
 part of $\Delta_\ep \Gatil$. The contributions arising from the 
 tree level transformation of $\Gatil_1$ are
\begin{eqnarray}
&& {3iN \over r^5} r_m (\ep \ga^m \theta)(\theta \thdot) \nn\\
&& + {3iA \over 2r^5}\left(2 r_m (\ep \ga^m \theta)(\theta \thdot)
- v_n (\ep \ga^m\theta)(\theta \ga^{mn}\theta) 
 - r_m (\ep \ga^n \thdot)(\theta \ga^{mn}\theta) \right)\nn\\
&& + {15 iA \over 4 r^7} r_m r_n v_l \left( (\ep \ga^n \theta)(\theta \ga^{ml}
\theta)+(\ep \ga^m \theta)(\theta \ga^{nl}
\theta)\right) \period \label{th3-01}
\end{eqnarray}
On the other hand, 
 the 1-loop level transformation applied to $\Gatil_0$ produces 
\begin{eqnarray}
-{3iN\over r^5} r_m (\ep \ga^m\theta)(\theta\thdot) \comma \label{th3-10}
\end{eqnarray}
which cancels the first term of (\ref{th3-01}). The remaining terms in 
(\ref{th3-01})  are all proportional to $A$. Now note that while
 the four-fermion structures in the second line of (\ref{th3-01}) 
have only one  \lq\lq free index", contracted with an arbitrary vector 
 $r_m(\tau)$ or $v_m(\tau)$, the ones in the last line 
 carry three free indices. Since the Fierz identities can only relate 
 structures with the same number of free indices, expressions in these two
 lines cannot cancel each other.  Moreover, it is easy to check, using the 
 Fierz identities given in the Appendix A, that the second line 
 does not vanish by itself. This then proves $A=0$.  

In summary, we have found that the  order 2 contribution at 1-loop for 
 the effective action in the background gauge is completely determined 
 by the requirement of supersymmetry and takes the form
\begin{eqnarray}
\Gatil_1 = \int\! d\tau\, N {\theta\thdot \over r^3} \period
\end{eqnarray}
This indeed agrees, including the overall normalization, 
 with the direct calculation performed in \cite{okawa9907}.
It can be easily checked that the normalization is directly linked to 
 the magnitude of the $\calO(\theta^0)$ quantum corrections in 
(\ref{delrfin}) and (\ref{delthfin}), which cannot be determined by 
 the closure property alone. 

As we shall discuss in the concluding section, the power of our 
 off-shell Ward identity can only be fully utilized at the next order in the 
 derivative expansion, where the most general form of the action unavoidably
 contains many terms with higher derivatives, such as $\ddot{r}_m$ and 
 $\ddot{\theta}_\al$. Nevertheless, it is gratifying that already at
 order 2 it has enabled us to see explicitly 
how the supersymmetry and the gauge symmetry intimately work together 
 to dictate the form of the effective action. 
\section{Summary and Discussions}
In this paper, we have derived the exact  supersymmetric  off-shell Ward 
 identity for Matrix theory 
as a step toward answering the \lq\lq old" yet important 
 unsettled  problem: \lq\lq To what extent do the symmetries, 
in particular the supersymmetry, 
  determine the low-energy effective action of Matrix theory?"
Our work was motivated by the observation that the existing analyses are 
incomplete in that off-shell trajectories with arbitrary 
time-dependence have not
 been fully considered. An important
 aspect of our Ward identity is that it allows the quantum-corrected 
effective supersymmetry transformation laws to be directly identified 
 in  closed form. They exhibit
 an intricate interplay with the gauge (BRST) symmetry of the theory, 
a feature not properly appreciated previously. 

As an application, we computed the explicit form of these 
 transformation laws at 1-loop to the lowest order in the derivative 
 expansion, and examined if the invariance under them determines 
the form of the effective action to the corresponding order. 
We found that the answer is affirmative,
 confirming the earlier result\cite{Pabanetal1}. This is as expected 
 since at this order the higher derivatives, such as the acceleration etc.,
can be eliminated from the effective action by  partial integration
 and the analysis is essentially the same as in the existing literature.  

The full significance of our off-shell Ward identity should become 
apparent starting from the next order, \ie from order 4, where 
complete elimination of higher derivatives will no longer be possible. 
There will be a considerable number of independent structures 
 allowed in the most general effective action. Even the proper 
 listing of them requires careful analysis
 due to the total derivative ambiguities  and the existence of non-trivial
 Fierz identities. Nonetheless, a preliminary investigation indicates that,
with an aid of computerized calculation, it appears feasible to determine 
 whether SUSY alone is enough to fix the form of the effective action at
 order 4 for arbitrary trajectories. 

Another important direction into which to extend our present work is 
 to apply our Ward identity to genuinely multi-body configurations. 
To find out 
 whether the remarkable agreement with supergravity in such a situation 
 \cite{Okawa-Yoneya} is due to supersymmetry alone would certainly 
 deepen our understanding of the Matrix theory further. 

At the more formal and structural level, we should mention that 
a further study should be made on the issue of 
 the closure  property of the effective SUSY 
transformations. 
As we have shown,  at the lowest order the closure turned out to be
 canonical. So far, however, we have not been able to answer whether 
 this persists at higher orders and loops. An analysis based on the 
 general closed form expressions for the transformation laws should 
 shed light on this intriguing question. 

We hope to be able to report on these and other related issues 
 in the near future. 
\par\bigskip\noindent
{\large\bf Acknowledgment}\par\smallskip\noindent
We would like to express our special gratitude to Y.~Okawa for 
 a number of clarifying discussions and his interest in our work. 
Y.~K acknowledges the warm hospitality extended to him 
by the organizers at the third international symposium 
on Frontiers of Fundamental Physics (Hyderabad, India), where a
preliminary version of this work was presented. 
This work is supported in part by Grant-in-Aid for Scientific Research
on Priority Area \#707 \lq\lq Supersymmetry and Unified Theory of 
 Elementary Particles", Grant-in-Aid for Scientific Research No.~09640337, 
 and Grant-in-Aid for International Scientific Research (Joint Research)
 No.~10044061,  from Japan Ministry of Education, Science and Culture. 
\newpage
\setcounter{equation}{0}
\renewcommand{\theequation}{A.\arabic{equation}}
\noindent
{\Large\bf Appendix A:} {\large\bf \quad Calculations of 
 $\calO(\theta^2)$ terms in $\Delta_\ep \theta_\al$}
\parbigskipn
In this appendix, we exhibit some details of the calculations of 
$\calO(\theta^2)$ terms in $\Delta_\ep \theta_\al$ at 1-loop order.

At this order, what we need to evaluate is (see (\ref{delth1}))
\begin{eqnarray}
\Delta_\ep \theta_{\al}(\tau) =
\Delta_\ep \theta^A_{\al}(\tau) + \Delta_\ep \theta^B_{\al}(\tau)
+\Delta_\ep \theta^C_{\al}(\tau)\comma 
\end{eqnarray}
where
\begin{eqnarray}
\Delta_\ep \theta^A_{\al}(\tau) &\equiv& 
 \langle \delta_\ep \psihat_{\al}(\tau)\rangle \comma \\
\Delta_\ep \theta^B_{\al}(\tau) &\equiv& 
 -\langle \delta_B \psihat_{\al}(\tau)\calO_\ep\rangle \comma \\
\Delta_\ep \theta^C_{\al}(\tau) 
&\equiv& - \int\! d\tau' 
\langle \delta_B \psihat_{\al}(\tau) \calO_m(\tau') \rangle
 {1\over i}(\ep\ga^m\theta)(\tau') \period
\end{eqnarray}
Hereafter in this appendix, $\Delta_\ep\theta_\al$ will refer only to the 
 $\calO(\theta^2)$ part. Also, we shall omit the overall factor of $N$,
 except in the final expression. The relevant Feynman diagrams are shown 
 in Fig.\ref{figure}.\\
\begin{figure}[h]
\begin{center}
\includegraphics[height=10.5cm]{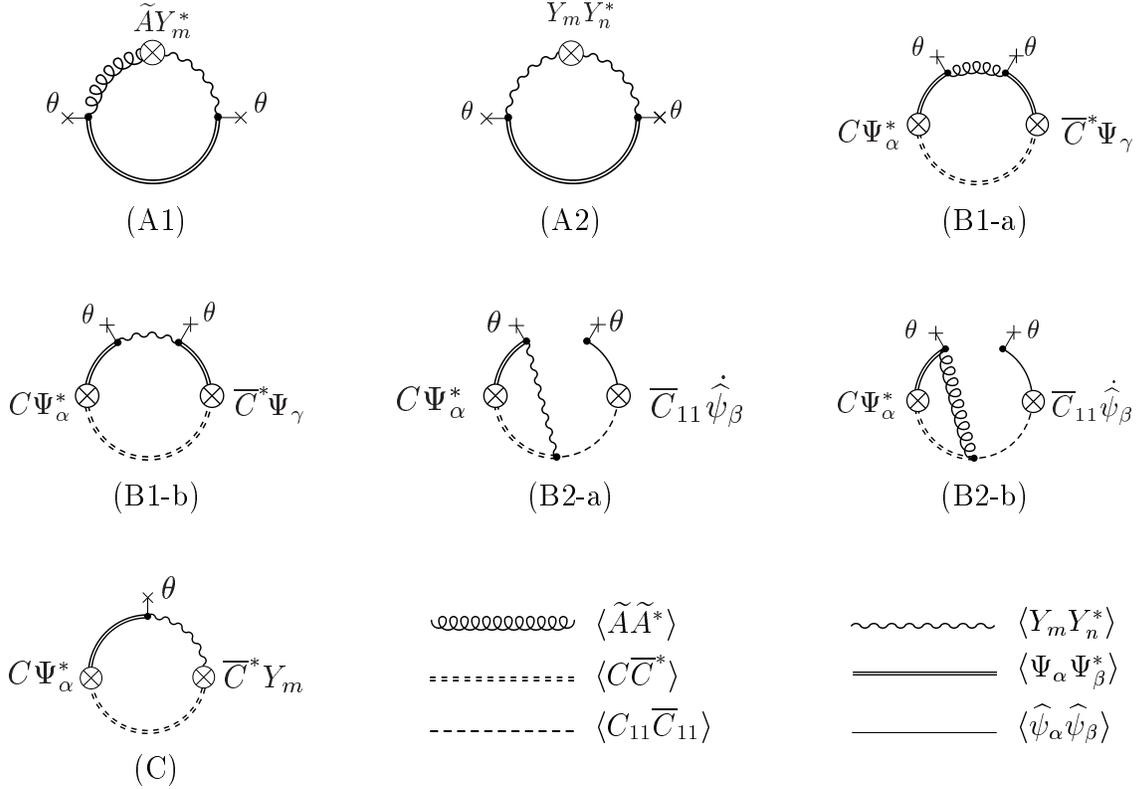}
\end{center}
\caption{Schematic depiction of Feynman diagrams relevant for the calculations 
 in this appendix.}
\label{figure}
\end{figure}
\parbigskipn
{\bf Calculation of $\Delta_\ep\theta^A_\al(\tau)$:}\qquad 
The explicit expression is 
\begin{eqnarray}
\Delta \theta^A_\al = g^2 \langle \Atil Y_m^\ast -Y_m \Atil^\ast\rangle
 (\ga^m \ep)_\al
+i{g^2\over 2}  \langle Y_m Y^\ast_n -Y_n Y^\ast_m\rangle
(\ga^{mn}\ep)_\al \period \label{deltha}
\end{eqnarray}
To compute the expectation values above to $\calO(\theta^2)$, 
we need to perform the second order perturbation using the 
$\calO(\theta)$ vertices
\begin{eqnarray}
\Stil_{\Atil \Psi} &=& -i \int\! d\tau \theta_\al(\Atil \Psi_\al^\ast
 -\Psi_\al \Atil^\ast) \comma \label{apsivertex}\\
\Stil_{Y\Psi} &=& -\int\! d\tau (\ga^m\theta)_\al (Y_m \Psi^\ast_\al -Y^\ast_m
 \Psi_\al)\period \label{ypsivertex}
\end{eqnarray}
This generates the diagrams (A1) and (A2) in Fig. \ref{figure}, 
for the first and the second term in (\ref{deltha}) respectively.
Consider for example the term 
 $\langle \Atil(\tau) Y_m^\ast(\tau)\rangle$. Using the tree-level
 propagators given in (\ref{PsiPsiast}) and (\ref{AAast}), this can be computed as 

\begin{eqnarray}
\langle \Atil(\tau) Y_m^\ast(\tau)\rangle &=& 
 i \int\! d\tau' d\tau'' \langle \Atil(\tau)\Atil^\ast(\tau')\rangle
 \theta_\al(\tau') (\ga^n\theta(\tau''))_\be 
\langle \Psi_\al(\tau')\Psi_\be^\ast(\tau'')\rangle \nn\\
&& \qquad \times \langle Y_n(\tau'')Y_m^\ast(\tau)\rangle\nn\\
&=& -i \bra{\tau} \Delta \theta_\al (\del+\rslash)_{\al\be} (\ga^m\theta)_\be 
 \Delta^2 \ket{\tau} \period
\end{eqnarray}
Performing the normal-ordering, neglecting the terms which generate
 derivatives, this becomes
\begin{eqnarray}
\langle \Atil(\tau) Y_m^\ast(\tau)\rangle = 
-i(\theta \rslash \ga^m\theta) \bra{\tau} \Delta^3 \ket{\tau}
= -{3i \over 16 r^5}(\theta \rslash \ga^m\theta)= {3i \over 16 r^5}
r_l (\theta \ga^{ml} \theta) \period
\end{eqnarray}
Likewise, one easily finds that $-\langle\Atil^\ast(\tau)
 Y_m (\tau)\rangle$ gives exactly the same contribution. The evaluation 
 of the diagram (A2) proceeds in an entirely similar manner. In this way 
 one finds
\begin{eqnarray}
\Delta \theta^A_\al = {3ig^2 \over 16 r^5}
\left(2 r_l (\theta \ga^{ml}\theta)
(\ga^m\ep)_\al
 -r_l(\theta \ga^{mnl}\theta) (\ga^{nm}\ep)_\al\right) \period
\end{eqnarray}
\parsmallskipn
{\bf Calculation of $\Delta_\ep\theta^B_\al(\tau)$:}\qquad 
This receives contributions from two classes of diagrams, (B1) and (B2)
 in Fig.\ref{figure}. 

For (B1), we have
\begin{eqnarray}
\Delta_\ep\theta^{B1}_\al(\tau)&=&
 \ep_\be g^2  \!\int\! d\tau' \Biggl(  \langle C^\ast(\tau)\Cbar(\tau')
\rangle(\del +\rslash)_{\be\ga} \langle \Psi^\ast_\ga (\tau')  
 \Psi_\al(\tau)\rangle \nn\\
&& -\langle C(\tau) \Cbar^\ast(\tau')\rangle
(\del -\rslash)_{\be\ga}\langle \Psi_\ga(\tau')
 \Psi^\ast_\al(\tau)\rangle \Biggr)\period
\end{eqnarray}
Insertions of the vertices $\Stil_{\Atil\Psi}$ (\ref{apsivertex})
 and $\Stil_{Y\Psi}$ (\ref{ypsivertex}) twice generate the diagrams 
 (B1-a) and (B1-b). Again neglecting the derivatives produced in the 
 process of normal ordering we obtain
\begin{eqnarray}
\Delta \theta^{B1}_\al 
={3ig^2\over 16 r^5}\left(2(\ep\theta) (\rslash \theta)_\al
-2(\ep\rslash\theta)\theta_\al-2r_l(\ep\ga^m\theta)(\theta\ga^{ml})_\al\right)
\period
\end{eqnarray}

Now, in distinction to all the other contributions, the one from (B2) involves 
propagation of massless fields $\hat{\psi}_\al$, $C_{11}$ and $\Cbar_{11}$
 in the intermediate steps. The original expression is, to the order of 
 interest, 
\begin{eqnarray}
\Delta_\ep\theta^{B2}_\al = 
-g \ep_\be \!\int\!d\tau^\prime
\langle C\Psi^\ast(\tau)\Cbar_{11}
\dot{\widehat{\psi}\ }\mbox{}\!\!_\beta(\tau^\prime)
\rangle  
+ g \ep_\be \!\int\!d\tau^\prime
\langle C^\ast\Psi(\tau)\Cbar_{11}
\dot{\widehat{\psi}}_\beta(\tau^\prime)\rangle \period
\end{eqnarray}
To extract $\calO(\theta^2)$ contributions, we need to use the following 
 five types of vertices:
\begin{eqnarray}
 V_1 &\equiv&  -ig\int\! d\tau \left(r^m\Cbar^\ast C_{11}Y^m +
 r^m \Cbar Y^{m\ast}C_{11}\right) \comma \\
 V_2 &\equiv& \int\! d\tau \theta \gamma^n
(\Psi Y^{m\ast} - Y^m \Psi^\ast) \comma  \\
 V_3 &\equiv& g\int\! d\tau(\Cbar^\ast \dot{C}_{11}\tilA +  
\Cbar^\ast C_{11}\dot{\tilA \ }\! 
- \Cbar\dot{\tilA^\ast} C_{11}
 -  \Cbar\tilA^\ast \dot{C}_{11} ) \comma  \\
 V_4 &\equiv& i\int\! d\tau \theta(\Psi\tilA^\ast - \tilA\Psi^\ast)  \comma \\
 V_5&\equiv& - \frac{1}{g^2}\int\! d\tau{\dot{\widehat{\psi}\ }}\theta \period
\end{eqnarray}
First, inserting $V_1, V_2$ and $V_5$, we get the diagrams of the type 
 (B2-a). This gives the contribution
\begin{eqnarray}
\Delta_\ep\theta^{B2-a}_\al = 
2ig^2\!\int\!d\tau^{\prime} \ 
d\tau^{\prime\prime}
\lk \tau | \Delta |\taupp\rk 
\lk \taupp|\frac{1}{\del}| \taup\rk
r^n(\taupp)\lk \taupp|\Delta
(\gamma^n\theta)_\alpha
\del\Delta|\tau \rk (\epsilon\theta)(\taup) \period\nn\\
\end{eqnarray}
Similarly, use of $V_3, V_4$ and $V_5$ generates the diagrams of the type
 (B2-b), the contribution of which is worked out to be
\begin{eqnarray}
\Delta_\ep\theta^{B2-b}_\al = 2ig^2 \langle \tau |
\Delta (\ep\theta) \Delta r^m (\ga^m\theta)_\al \Delta |\tau\rangle 
 - \Delta_\ep\theta^{B2-a}_\al \period
\end{eqnarray}
Therefore, 
\begin{eqnarray}
\Delta_\ep \theta^{B2}_\al = \Delta_\ep \theta^{B2-a}_\al
+\Delta_\ep \theta^{B2-b}_\al = 2ig^2(\ep\theta) \langle \tau |\Delta^3|
\tau\rangle (\rslash\theta)_\al
 = {3ig^2 \over 16 r^5} 2 (\ep\theta) (\rslash\theta)_\al \period
\end{eqnarray}
\parsmallskipn
{\bf Calculation of $\Delta_\ep\theta^C_\al(\tau)$:}\qquad 
Finally, consider $\Delta_\ep\theta^C_\al$. It takes the form 
\begin{eqnarray}
\Delta_\ep\theta^C_\al(\tau) &=& 
g^2 \!\int\! d\tau' \Biggl( \langle C(\tau)\Cbar^\ast(\tau')\rangle 
 \langle Y_m(\tau') \Psi_\al^\ast(\tau)\rangle \nn\\
&& \qquad + \langle C^\ast(\tau) \Cbar(\tau')\rangle 
\langle Y^\ast_m(\tau') \Psi_\al(\tau)\rangle \Biggr) \ep^T\ga^m\theta(\tau')
\comma 
\end{eqnarray}
which is represented by the diagram $(C)$. 
Using the vertex (\ref{ypsivertex}) and proceeding similarly to 
 the previous calculations, we obtain
\begin{eqnarray}
\Delta \theta^C_\al ={3ig^2 \over 16r^5}\left(-2
(\ep\rslash\theta)\theta_\al-2r_l(\ep\ga^m\theta)(\theta\ga^{ml})_\al\right)
\period
\end{eqnarray}
\parsmallskipn
{\bf Summary:} \qquad Adding up all the contributions and 
 reinstating the factor of $N$, the final result is
\begin{eqnarray}
\Delta_\ep\theta_\al &=& 
{3ig^2N \over 16 r^5} \Biggl(
-r_l (\theta\ga^{mnl}\theta) (\ga^{nm}\ep)_\al
 + 2r_l(\theta \ga^{ml}\theta) (\ga^m\ep)_\al -4r_l(\ep\ga^m\theta)
(\theta \ga^{ml})_\al \nn\\
&& \qquad +4(\ep\theta)(\rslash\theta)_\al 
 -4(\ep \rslash \theta)\theta_\al\Biggr) \period 
\end{eqnarray}
\setcounter{equation}{0}
\renewcommand{\theequation}{B.\arabic{equation}}
\parbigskipn
{\Large\bf Appendix B:} {\large\bf \quad $SO(9)$ Fierz Identities}
\parbigskipn
In this appendix, we record the $SO(9)$ Fierz identities which are
 crucial in simplifying the $\calO(\theta^2)$ part of $\Delta_\ep \theta_\al$
 at 1-loop. 

Adapting the notations of Taylor and Raamsdonk\cite{TR}, let
 us introduce the following quantities for 
 $n=0,1,2,3$ (repeated indices are summed):
\begin{eqnarray}
E_n &=& r_m(\ep \ga^{a_1 \cdots a_n m} \theta)
(\theta \ga^{a_n \cdots a_1}\lam)\comma 
\\
\Ebar_n &=& r_m(\ep \ga^{a_1 \cdots a_n } \theta)(\theta \ga^{a_n \cdots a_1m}
\lam)\comma  \\
F_n &=& r_m(\theta \ga^{a_1 \cdots a_n m} \theta)(\lam \ga^{a_n \cdots a_1}
\ep)\comma  \\
\Fbar_n &=& r_m(\theta \ga^{a_1 \cdots a_n } \theta)
(\lam \ga^{a_n \cdots a_1m}\ep)\comma  \\
P &=& {1\over 4!} r_m \ep^{a_1a_2a_3a_4 b_1b_2b_3b_4 m}
(\ep \ga^{a_1a_2a_3a_4}\theta)( \theta \ga^{b_1b_2b_3b_4 m}\lam)\comma \\
Q &=& {1\over 4!} r_m \ep^{a_1a_2a_3a_4 b_1b_2b_3b_4 m}
(\theta \ga^{a_1a_2a_3a_4}\theta)( \lam \ga^{b_1b_2b_3b_4 m}\ep) \comma 
\end{eqnarray}
where $\lam$ is an arbitrary spinor. 
Because  $\theta \Ga \theta =0$ for any
 symmetric matrix $\Ga$,  five of them  actually vanish, namely
\begin{eqnarray}
F_0=\Fbar_0 = \Fbar_1 = F_3 = Q=0 \period
\end{eqnarray}
Since there are nine independent Fierz identities\cite{TR}, only four 
structures are independent, which we take to be $F_1, F_2,
 \Fbar_2$ and $\Fbar_3$. Then the remaining quantities can  be expressed 
 in terms of them as\footnote{The sign in front of $48E_0^-$ in 
Eq.~(B.20) of \cite{TR} should be $+$. }
\begin{eqnarray}
E_0 &=& {1\over 16}F_1-{1\over 32} (F_2+\Fbar_2) -{1\over 96} \Fbar_3 \comma \\
\Ebar_0 &=& -{1\over 16}F_1-{1\over 32}(F_2+\Fbar_2) +{1\over 96} \Fbar_3 
\comma \\
E_1 &=& -{3\over 8}F_1-{1\over 8}(F_2-\Fbar_2) -{1\over 48}\Fbar_3 \comma \\
\Ebar_1 &=& -{3\over 8}F_1+{1\over 8}(F_2-\Fbar_2) -{1\over 48}\Fbar_3 
\comma \\
E_2 &=& {7\over 4}F_1 -{1\over 4}(F_2+\Fbar_2) +{1\over 24} \Fbar_3 \comma \\
\Ebar_2 &=& -{7\over 4}F_1 -{1\over 4}(F_2+\Fbar_2) -{1\over 24} \Fbar_3 
\comma \\
E_3 &=& -{21\over 4}F_1 +{3\over 4}(F_2-\Fbar_2) +{3\over 8} \Fbar_3 \comma \\
\Ebar_3 &=& -{21\over 4}F_1 -{3\over 4}(F_2-\Fbar_2) +{3\over 8} \Fbar_3 
\comma \\
P &=& {15\over 2} (F_2+\Fbar_2) \period
\end{eqnarray}
>From these relations, one easily finds the identity
\begin{eqnarray}
0= F_2-2F_1  -4\Ebar_1 +4\Ebar_0+12 E_0 \period \nn
\end{eqnarray}
Removing $\lam$, we get 
\begin{eqnarray}
0&=&  -r_l(\theta \ga^{mnl}\theta)(\ga^{nm}\ep)_\al
 +2 r_l(\theta \ga^{nl}\theta)( \ga^{n}\ep)_\al
+4  r_l(\ep \ga^n \theta)(\ga^{nl}\theta)_\al \nn\\
&& \qquad +4r_l(\ep\theta)(\ga^l\theta)_\al + 12 r_l(\ep \ga^{l}\theta)
\theta_\al \comma 
\end{eqnarray}
which was used in Sec.~5. 

\end{document}